\documentclass[]{raa}            
\usepackage{graphicx,times}
\usepackage{natbib}

\providecommand{\bm}[1]{\mbox{\boldmath$#1$\unboldmath}}

\begin{document}

\title{Steady State Equilibrium Condition of $npe^{\pm}$ Gas and Its Application to Astrophysics
$^*$ \footnotetext{\small $*$Partially supported by the National
Natural Science Foundation of China (10733010,10673010,10573016),
National Basic Research Program of China (2009CB824800), Youth Fund
of Sichuan Provincial Education Department (2007ZB090,2009ZB087) and
Science and Technological Foundation of CWNU(09A004)}}

 \volnopage{ {\bf 20xx} Vol.\ {\bf 9} No. {\bf XX}, 000--000}
   \setcounter{page}{1}

   \author{Men-Quan Liu
      \inst{1,2}
        }

   \institute{Center for Astrophysics, University of Science and
Technology of China, Hefei Anhui, 230026, P.R. China; {\it menquan@mail.ustc.edu.cn}\\
        \and    Institute of Theoretical Physics, China West Normal
University, Nanchong Sichuan, 637002, P.R. China
         \no\\
   {\small Received [year] [month] [day]; accepted [year] [month] [day] }
}

\abstract{The steady equilibrium conditions for a mixed gas of
neutrons, protons, electrons, positrons and radiation field
(abbreviated as $npe^{\pm}$ gas) with/without external neutrino flux
are investigated, and a general chemical potential equilibrium
equation $\mu_n=\mu_p+C\mu_e$ is obtained to describe the steady
equilibrium at high temperatures ($T>10^9$K). An analytic fitting
formula of coefficient $C$ is presented for the sake of simplicity
as the neutrino and antineutrino are transparent. It is a simple
method to estimate the electron fraction for the steady equilibrium
$npe^{\pm}$  gas that using the corresponding equilibrium condition.
As an example, we apply this method to the GRB accretion disk and
approve the composition in the inner region is approximate
equilibrium as the accretion rate is low. For the case with external
neutrino flux, we calculate the initial electron fraction of
neutrino-driven wind from proto-neutron star model M15-l1-r1. The
results show that the improved equilibrium condition makes the
electron fraction decrease significantly than the case
$\mu_n=\mu_p+\mu_e$ when the time is less than 5 seconds post
bounce, which may be useful for the r-process nucleosynthesis.
 }

   \authorrunning{M.-Q. Liu}            
   \titlerunning{Steady State Equilibrium Condition of $npe^{\pm}$ Gas }  
   \maketitle


%
%

\section{Introduction}
It is a classic and simple approximation for the practical
application at many astrophysical sites that matter compositions can
be considered as a mixture of the neutrons, protons and electrons,
i.e. so called $npe^-$ system. If the temperature is very
high($T>10^9$K), lots of photons, positrons, even neutrinos and
antineutrinos will appear in the system, i.e., the system becomes a
mixture of electrons, positrons, nucleons and radiation field (we
abbreviated it as $npe^{\pm}$ gas). Many astrophysical sites can be
regarded as the $npe^{\pm}$ gas, such as (i) the hot fireball jetted
from a successful central engine of Gamma Ray Burst (GRB)
\citep{Pruet2002}, (ii) the matter after the core-collapse supernova
shock due to the photodisintegration of the iron
nuclei\citep{Marek2009}, (iii)the neutrino-driven wind comes from
the proto-neutron star(PNS) as $T>10^9$K \citep{Martinez2008},(iv)
the outer core of the young neutron
star\citep{Yakovlev2008,Baldo2009}, (v)the accreting disk of the
GRBs \citep{Liu2007,Janiuk2010} and (vi) the early universe before
the decoupling of neutrinos \citep{Dutta2004,Harwit2006}. In a word,
$npe^-$ and $npe^{\pm}$ gas is applied widely up to the present.
Steady equilibrium state of $npe^-$ or $npe^{\pm}$ gas is an
important stage for many cases. Many authors have addressed this
issue for several decades. A typical disposal to the steady sate
equilibrium of $npe^-$ system is concluded by \citet{Shapiro1983}.
They gave an important result that $\mu_n=\mu_p+\mu_e$ for a steady
equilibrium $npe^-$ system, where $\mu^{'}s$ are the chemical
potentials for neutron,proton and electron respectively. This result
have been accepted by most authors. But in fact, Shapiro et al. only
considered the electron capture and its reverse interaction at 'low
temperature', they ignored the appearance of positrons when the
temperature of system is high enough. Recently,\citet{Yuan2005}
argued that lots of positions can exit at high temperature, which
leads to the great increase of the position capture rate. The
positron capture affects the condition of steady equilibrium
significantly. If the neutrinos can escape freely from the system
with plenty of $e^{\pm}$ pairs, the equilibrium condition should be
$\mu_n=\mu_p+2\mu_e$ instead. However, for a more general condition
when the temperature is moderate, the equilibrium condition have not
be researched. Liu et al. have ever taken a method in which they
assume the coefficient of $\mu_e$ varies exponentially from
$\mu_n=\mu_p+\mu_e$ to $\mu_n=\mu_p+2\mu_e$ in the accreting disk of
GRBs\citep{Liu2007}, but it is not a rigorous method. Therefore a
detailed and reliable database or fitting function to describe
steady equilibrium of the $npe^{\pm}$ gas at any temperature is
necessary. Furthermore, the above discussions  are limited to the
isolated system, ignoring the external neutrino flux. In this paper
we investigate the chemical equilibrium condition for $npe^{\pm}$
gas at any temperature from $10^9$ to $10^{11}$K, and give a
concrete application to the GRB accretion disk. We also calculate
the initial electron fraction of the neutrino-driven wind in PNS, in
which the external strong neutrino flux can not be ignored. This
paper is organized as follows. In section II, we present the
equilibrium conditions as neutrino is transparent or opaque for an
isolated system. Section III contains a detailed discussion to the
initial electron fraction of neutrino-driven wind for PNS model
M15-l1-r1. Finally we analyze the results and make our conclusions.

\section{Equilibrium Condition of $npe^{\pm}$ Gas without the external neutrino flux}
For a mixed gas of $npe^{\pm}$ and radiation field at different
physical conditions, we divide them into two cases: neutrino
transparence and opacity. To guarantee the self-consistence, we give
a simple estimate for the opaque critical density of the $npe^{\pm}$
gas. The mean free path of neutrino is
$l_{\nu}=\frac{1}{n\sigma^{sac}_{\nu}+n_n\sigma^{abs}_{\nu}}$, where
$n$ and $n_n$ are the number density of baryon and neutron
respectively. $n=\rho N_A$, $n_n=\rho(1-Y_e)N_A$, $\rho $ is the
mass density, $Y_e$ is the electron fraction, $N_A$ is the
Avogadro's constant, $\sigma^{sac}_{\nu}$ and $\sigma^{abs}_{\nu}$
are the scatter cross section with baryons and absorption  section
by the neutrons.
$\sigma^{sac}_{\nu}\approx(\frac{E_{\nu}}{m_ec^2})^210^{-44}$\citep{Kippenhanhn1990},
where $E_{\nu}$ is the energy of neutrino, $m_e c^2$ is the mass
energy of electron, $c$ is the light velocity.
$\sigma^{abs}_{\nu}\approx\frac{A}{\pi^2}E_ep_e\approx\frac{A}{\pi^2}E^2_e
$\citep{Qian1996,Lai1998}, where $A=\pi
G^2_Fcos^2\theta_c(C^2_V+3C^2_A)$, $G_F=1.436\times 10^{-49}$ erg
cm$^3$ is the Fermi weak interaction constant, cos$^2\theta_c=0.95$
refers to Cabbibo angle. $C_V=1$, $C_A=1.26$, $E_e$ and $p_e$ are
the energy and momentum for electron, respectively. Due to the
energy conversation of nuclear reaction, $E_e=E_{\nu}+Q$,
$Q=(m_n-m_p)c^2=1.29$MeV, $m_n$ and $m_p$ are the mass of neutron
and proton. At high density, the electrons are strong degenerate and
relativistic, so $E_e\approx
E_F=[(3\pi^2\bar{\lambda}^3_en_e)^{2/3}+1]^{1/2}$(in unite of
$m_ec^2$), $\bar{\lambda}_e=\frac{\hbar}{m_ec}$ is the reduced
electron Compton wavelength. Substituting $\rho Y_e N_A$ for $n_e$,
$E_e\approx(3\pi^2\bar{\lambda}^3_e\rho Y_eN_A)^{1/3}$. Therefore,
the mean free path of neutrino is
\begin{equation}
l_{\nu}=\frac{1}{\rho N_A[(3\pi^2\bar{\lambda}^3_e\rho
Y_eN_A-Q)^{2/3}\times10^{-44}]+[\frac{A}{\pi^2}(3\pi^2\bar{\lambda}^3_e\rho
Y_eN_A)^{2/3}]\rho(1-Y_e)N_A}.
\end{equation}
If we assume $l_{\nu}=10 \rm km$ is criterion of neutrino opacity,
$\rho^{\nu}_{cri}=5.58,4.50,4.10,3.96,3.96\times 10^{10} $g
cm$^{-3}$ for $Y_e=0.1,0.2,0.3,0.4,0.5$. Rigorously speaking, here
we  overestimate the absorption section because we ignore a block
factor $(1-f_e)$, so $\rho^{\nu}_{cri}$ is the minimum for critical
density. As $\rho<\rho^{\nu}_{cri}$, neutrino is transparent, or it
is opaque. By similar method, one only needs to replace $\nu$,
$E_e=E_{\nu}+Q$, and $n_n$ to $\bar{\nu}$, $E_e=E_{\bar{\nu}}-Q$,
and $n_p$ respectively for mean free path of antineutrino. The
critical density for antineutrino
$\rho^{\bar{\nu}}_{cri}=1.43,0.86,0.62,0.48,0.40\times 10^{11} $g
cm$^{-3}$ for $Y_e=0.1,0.2,0.3,0.4,0.5$.

Another preciser way to judge the transparency of neutrino is
defining a parameter: neutrino optical depth $\tau$, which is
closely relative to object's composition and structure. $\tau=\int
^\infty_r <\kappa_{eff}>dr$\citep{Arcones2008}, where $r$ is
neutrino transport distance, $
<\kappa_{eff}>=\sqrt{<\kappa_{abs}>(<\kappa_{abs}>+<\kappa_{sac}>)}$,
$\kappa_{abs}$ and $\kappa_{sac}$ are the absorption opacity and
scatter opacity, $\kappa_{sac}=n\sigma_{sac}$,
$\kappa_{abs}=\sum_i{n_i\sigma_{abs(i)}}$, $\sigma_{abs(i)}$ and
$n_i$ is neutrino absorption cross section and number density of
target particle. Usually authors define $\tau<\frac{2}{3}$ or 1 as
the criterion for neutrino transparent\citep{Cheng2009,Janka2001}.
Following we investigate chemical equilibrium condition for two
different cases respectively.

 \subsection{Case 1. Neutrinos are Transparent}
 When the $npe^{\pm}$ gas  is in equilibrium
 and transparent to neutrino and antineutrino ($\mu_{\nu}=\mu_{\bar{\nu}}=0,
$ that is, we have taken their number densities to be zero), the
beta reactions are the most important physical
processes\citep{Yuan2005}.  The steady equilibrium state is achieved
via the following beta reactions, \begin{eqnarray}
e^-+p &\rightarrow & n+ \nu_e, \label{2}\\
e^++n &\rightarrow & p+ \bar{\nu}_e, \label{3}\\
n&\rightarrow & p+e^- + \bar{\nu}_e .  \label{4}
 \end{eqnarray}
Reactions(\ref{2})-(\ref{4}) denote the electron capture(EC),
positron capture(PC) and beta decay(BD) respectively. Since the
system is transparent to neutrino and antineutrino, neutrinos and
antineutrinos  produced by reaction(\ref{2})-(\ref{4}) can escape
freely at once, inducing lots of energy loss, so the reverse
reactions, neutrino capture and antineutrino capture, are
negligible. If the system is in equilibrium sate, composition is
fixed and electron fraction $Y_e$ keeps as  constant. EC decreases
the $Y_e$, while PC and BD increase $Y_e$, then a general steady
equilibrium condition is
\begin{equation}
 \lambda_{e^{-}p}=\lambda_{e^+n}+\lambda_{n},
 \label{5}
 \end{equation}
where $\lambda^{'}$s are the reaction rates, subscript symbols
denote reaction particles (the same in following section). Other
reactions such as $\gamma+\gamma \leftrightarrow e^-
+e^+\leftrightarrow \nu +\bar{\nu}$ also exist, but they do not
influence the electron fraction directly. These beta reaction rates
can be obtained
 in the previous studies, we list them as below(We here employ the
natural system of units with $m_e=\hbar = c = 1$. In normal units,
they would be multiplied by $\frac{(m_ec^2)^5c}{(\hbar c)^7}$)
\citep{Yuan2005,Langanke2000},
 \begin{eqnarray}
\lambda_{e^-p}&\simeq&\frac{A}{2\pi^4}n_p\int_Q^{\infty}dE_e
        E_ep_e(E_e-Q)^2F(Z,E_e)f_e,
        \label{eq:rate_e_cap2} \\
\lambda_{e^+n}&\simeq&\frac{A}{2\pi^4} n_n\int_{m_e}^{\infty}dE_e
        E_ep_e(E_e+Q)^2F(-Z,E_e)f_{e^+},
        \label{eq:rate_p_cap2} \\
\lambda_{n}&\simeq&\frac{A}{2\pi^4}n_n\int_{m_e}^QdE_e
        E_ep_e(Q-E_e)^2F(Z+1,E_e)(1-f_{e}),
        \label{eq:rate_n_dec2}
\end{eqnarray}
Considering charge neutrality, $Y_e = Y_p$, and  the conservation of
the baryon number, $Y_n+Y_p=1$, so $n_p$ and $n_n$ in Eq.s (6)-(8)\,
are equal to $\rho Y_e N_A$ and $\rho(1-Y_e)N_A$, respectively.
$F(\pm Z,E_e)$ is Fermi function, which corrects the phase space
integral for the Coulomb distortion of the electron or positron wave
function near the nucleus. It can be approximated by
\begin{equation}
F(\pm Z,E_e)\approx 2(1+s)(2p_eR)^{2(s-1)}e^{\pi \eta}|\frac{\Gamma
(s+i\eta)}{\Gamma (2s+1)}|^2,
\end{equation}
here $Z$ is the nuclear charge of the parent nucleus, $Z=1/0$ for
proton/neutron, $s=(1-\alpha ^2 Z^2)^{1/2}$, $\alpha$ is the fine
structure constant, $R$ is the nucleus radius, $\eta=\pm \alpha Z
E_e /p_e$, $\Gamma(x)$ is the Gamma function. We do not adopt any
limiting form for Fermi function. Comparing to the derivation of the
rates in Yuan(2005), we consider additionally the Coulomb screening
of the nuclei. $ f_e$ and $f_{e^+}$ are the Fermi-Dirac functions
for electron and positron.
$f_e=[1+\exp{(\frac{E_e-\mu_e}{kT})}]^{-1}$,
$f_{e^+}=[1+\exp{(\frac{E_e+\mu_e}{kT})}]^{-1}$, where  $k$ is the
Boltzmann's constant, electron chemical potential $\mu_e$ can be
calculated as follows(energy is in unit of $m_ec^2$ and momentum in
unit of $m_ec$),
 \begin{equation}
\rho N_A Y_e=\frac{8\pi}{\lambda_{e}^3
}\int_0^{\infty}(f_{e}-f_{e^+})p^2dp.
\end{equation}
where $\lambda_{e}=\frac{h}{m_ec}$ is the electron Compton
wavelength. Note that the calculation method of chemical potential
of electron (including the chemical potentials of proton and neutron
in Eqs.(11)-(12)) also differs from the method in Yuan (2005). For
one system with the given temperature $T$ and densities $\rho$,
electron fraction $Y_e$ can be determined by iteration technique of
Eq.(\ref{5}).

 Figure 1
shows the  $T$, $Y_e$ and $\rho$  that satisfy the equilibrium
condition. It can be found that the $Y_e$ decrease with the
densities. As $\rho>10^{11} \rm g \, cm^{-3}$, $Y_e$ tends to zero,
especially for the lower temperatures. This consists with the
results in Fig. 5 of Ref. \citep{Reddy1998}. At the high density,
the $\beta$ decay is almost forbidden and the positron capture rate
is smaller than that of electron. In order to sustain the
equilibrium, electron number density $n_e$ must be very low, which
causes the $Y_e$ decrease obviously. Note that it quite differs from
the direct Urca process for strong degenerate
baryons\citep{Shapiro1983}, in which $n_p/n_n>1/8$. The baryons here
are nondegenerate since their chemical potentials (minus their rest
mass) are very low, even minus. After $\rho,T$ and $Y_e$ are found,
chemical potentials $\mu_{n},\mu_{p}$ can be calculated as
below(energy is in unit of $m_ec^2$ and momentum in unit of $m_ec$),
 \begin{eqnarray}
 \rho N_AY_p=\frac{8\pi}{\lambda_{e}^3}
\int_0^{\infty}p^2[1+\exp{(\frac{E_p-\mu_p}{kT})}]^{-1}dp, \\
\rho N_A(1-Y_e)=\frac{8\pi}{\lambda_{e}^3}
\int_0^{\infty}p^2[1+\exp{(\frac{E_n-\mu_n}{kT})}]^{-1}dp,\
 \end{eqnarray}
 where the conservation of the baryon number and the charge density are also included.
\begin{table*}
\centering
 \caption{\small\label{Tab1}steady state chemical equilibrium condition as neutrino is transparent for $T=10^9\rm K$.
 $\mu_p^{'}$\, $\mu_e^{'}$ and $\mu_n^{'}$ are chemical potentials without rest mass.}
  \vspace{12pt}
\begin{tabular}{c c c c c c c c c}
\hline \hline $Y_e$&$\rho$&$\lambda_{e^-p}$&$\lambda_{e^+n}$&$\lambda_{n}$&$\mu_e^{'}$&$\mu_n^{'}$&$\mu_p^{'}$&$C$\\
$$&$\rm g\,cm^{-3}$&$\rm cm^{-3}\,s^{-1}$&$\rm cm^{-3}\,s^{-1}$&$\rm cm^{-3}\,s^{-1}$&$\rm MeV$&$\rm MeV$&$\rm MeV$&$$\\
\hline
0.10     &   1.50E+08    &   8.56E+26    &   3.52E+19    &   8.56E+26  &   0.84    &   -0.43  &   -0.62  &   1.09    \\
0.20    &   6.83E+07    &   5.21E+26    &   2.22E+19    &   5.21E+26   &   0.80    &   -0.51  &   -0.63  &   1.07    \\
0.33     &   4.27E+07    &   3.72E+26    &   1.63E+19    &   3.72E+26   &   0.78   &   -0.56  &   -0.63  &   1.06    \\
0.40     &   3.02E+07    &   2.79E+26    &   1.26E+19    &   2.79E+26   &   0.76   &   -0.60  &   -0.64  &   1.05    \\
0.50   &   2.29E+07    &   2.14E+26    &   9.98E+18    &   2.14E+26     &   0.74  &   -0.64  &   -0.64  &   1.03    \\
\hline
\end{tabular}
\end{table*}

\begin{table*}
\centering
 \caption{\small\label{Tab2}steady state chemical equilibrium condition as neutrino  is transparent for $T=5\times 10^9\rm K $.
 The notes are the same as those in Table 1.}
  \vspace{12pt}
\begin{tabular}{c c c c c c c c c}
\hline \hline $Y_e$&$\rho$&$\lambda_{e^-p}$&$\lambda_{e^+n}$&$\lambda_{n}$&$\mu_e^{'}$&$\mu_n^{'}$&$\mu_p^{'}$&$C$\\
$$&$\rm g\,cm^{-3}$&$\rm cm^{-3}\,s^{-1}$&$\rm cm^{-3}\,s^{-1}$&$\rm cm^{-3}\,s^{-1}$&$\rm MeV$&$\rm MeV$&$\rm MeV$&$$\\
\hline
0.10     &   2.32E+08    &   2.00E+29    &   1.57E+29    &   4.28E+28  &   0.64     &   -3.00   &   -3.94   &   1.95    \\
0.20     &   8.38E+07    &   9.51E+28    &   7.73E+28    &   1.79E+28   &   0.45    &   -3.49   &   -4.08   &   1.96    \\
0.30    &   4.43E+07    &   5.71E+28    &   4.75E+28    &   9.61E+27     &   0.33   &   -3.82   &   -4.18   &   1.97    \\
0.40   &   2.71E+07    &   3.71E+28    &   3.14E+28    &   5.64E+27     &   0.23    &   -4.10   &   -4.27   &   1.98    \\
0.50      &   1.78E+07    &   2.47E+28    &   2.13E+28    &   3.38E+27    &   0.14  &   -4.35   &   -4.35   &   1.99    \\
\hline
\end{tabular}
\end{table*}

In order to describe the numerical relationship of $\mu_e$, $\mu_n$
and $\mu_p$, we define a factor $C$: $\mu_n=\mu_p+C\mu_e$. Table 1
and Table 2 are the results at $T=10^9$K and $5\times 10^9$K
respectively. It can be seen from Table 1 that
$\lambda_{e^-p}\approx \lambda_{n}>>\lambda_{e^+n}$ , i.e., positron
capture rate at this time can be ignored. $C \approx 1$ means that
$\mu_n=\mu_p+\mu_e$ is valid. While from Table 2 one can find
$\lambda_{e^-p}\approx \lambda_{e^+n}>>\lambda_{n}$, i.e., beta
decay becomes neglectable because lots of positrons take part in the
reactions at high temperature. Correspondingly, the equilibrium
condition becomes to $\mu_n=\mu_p+C\mu_e, C\approx2$. It is quite
different to the well known result $\mu_n=\mu_p+\mu_e$. This result
was first observed by Yuan (2005), and a detailed  explanation can
be found in\citep{Yuan2005}. Here we give a simple explanation that,
$\lambda_{e^-p}\propto n_en_p\propto f_e f_p$,
$\lambda_{e^+n}\propto
 n_{e^+}n_n\propto f_nf_{e^+}$, so $\lambda_{e^-p}-\lambda_{e^+n}\propto f_e f_p -
f_nf_{e^+}=f_pf_{e^+}(\frac{f_e}{f_{e^+}}-\frac{f_n}{f_p})$.
Considering $f_e\approx \exp{(\frac{E_e-\mu_e}{kT})}$,
$f_{e^+}\approx \exp{(\frac{E_e+\mu_e}{kT})}$, $f_p\approx
\exp{(\frac{E_p-\mu_p}{kT})}$ and $f_n\approx
\exp{(\frac{E_n-\mu_n}{kT})}$, we find $\mu_n=\mu_p+2\mu_e$ is valid
as $\lambda_{e^-p}=\lambda_{e^+n}$. For a more universal case, none
of $\lambda_{e^-p}$, $\lambda_{e^+n}$ and $\lambda_{n}$ can be
ignored, the coefficient $C$ will vary with the physical conditions.
Figure 2 shows the coefficient $C$ at different $T$ and $Y_e$. It
can be found that $C$ mainly depends on temperature $T$. When
$T<10^9\rm K$, $C\approx 1$; when $T$ from $10^9\rm K$ increases to
$5\times 10^9 \rm K$, $C$ increases significantly from 1 to 2;  when
$T>5\times 10^9 \rm K$, $C\approx 2$. But when $T>3\times 10^{10}$K
and $Y_e
>0.4$, $C$ is obviously larger than 2. The reason is that the
fiducial analysis in reference \citep{Yuan2005} ignoring  the Fermi
function. If we set Fermi functions are equal to 1, $C\approx 2$ is
still valid. For the convenience to practical application, we give
an analytic fitting formula that can facilitate application,
\begin{equation}
C=2-[1+\exp{(\frac{T_9-A_i}{B_i})}]^{-1}, \label{fitting}
\end{equation}
where $A=[2.8643,2.9249,2.9785,2.9902,3.0094],B=[0.79138,0.72181,
0.66331, 0.61813,0.57999]$ corresponding to $Y_e
=[0.1,0.2,0.3,0.4,0.5]$. $T_9$ is the temperature in unites of
$10^9\,\rm K$ ($T_9\in[1-6]$). The accuracy of the fitting is
generally better than 1\%.

\begin{figure*}[htb]
\centering
\includegraphics{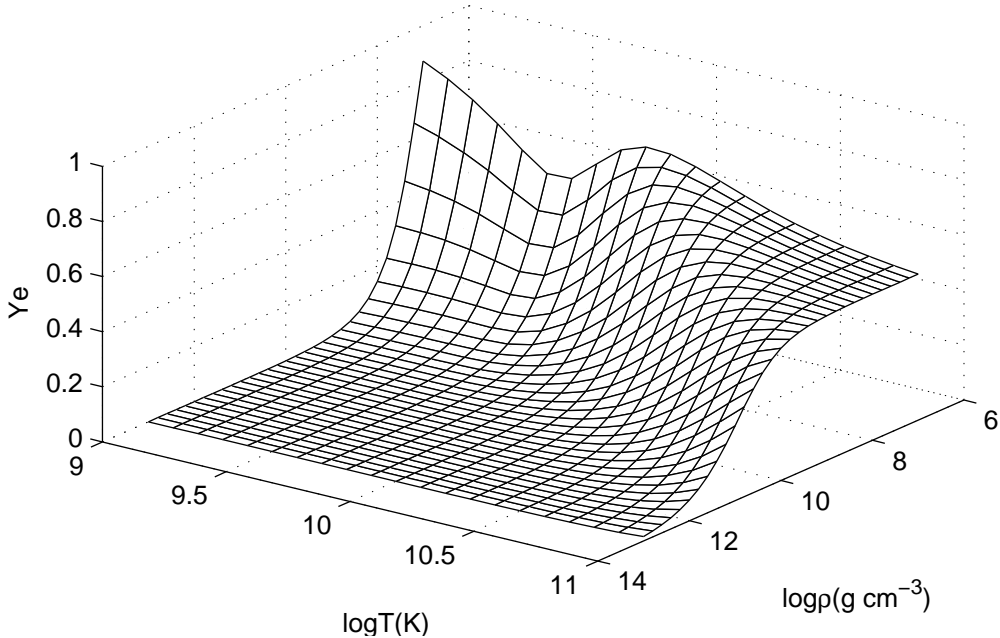} \caption{\label{Fig2} electron fraction $Y_e$ as a function of $T$ and $\rho$ for equilibrium state $npe^{\pm}$ gas.}
\end{figure*}

\begin{figure*}[htb]
\centering
\includegraphics{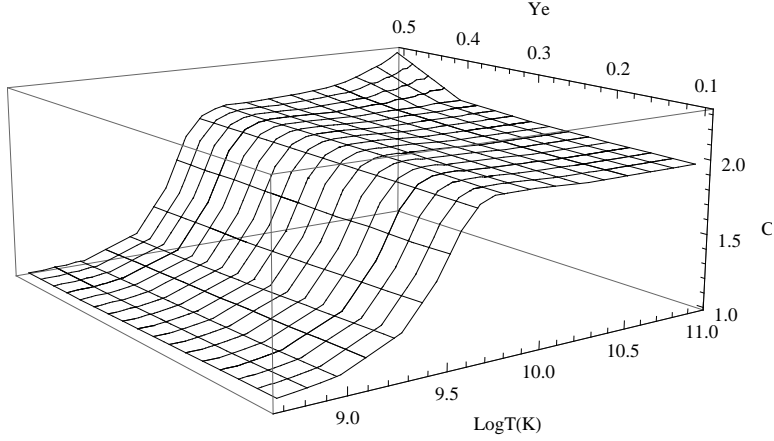} \caption{\label{Fig1} the coefficient  $C$ of chemical potential equilibrium condition $\mu_n=\mu_p+C\mu_e$ as a function of $T$ and $Y_e$.}
\end{figure*}

As an example, we introduce the application to electron fraction of
 GRB accretion disk. GRB is one of the most violent events in the
universe, but its explosion mechanism is still not clear. Many
authors support the view that GRB originates from the accretion disk
of stellar mass black hole. Various accretion rates (from
$0.01M_{\odot} \, s^{-1}$ to 10$M_{\odot}\,s^{-1}$) bring quite
significant difference to the disk structure and composition. As the
temperature of the accretion disk is generally  larger than
$10^{10}K$, all nuclei are dissociated to the free nucleons, so
$npe^{\pm}$ gas can describe the composition well. For lower
accretion rates ($\dot{M}\leq0.1M_{\odot}\,s^{-1}$), the disk is
transparent to neutrinos and antineutrinos, and neutrino and
antineutrino absorption are not important\citep{Surman2004}.
Adopting the steady equilibrium condition,  $Y_e$  of the disk model
PWF99\citep{Popham1999}(accretion rate
$\dot{M}=0.1M_{\odot}\,s^{-1}$, alpha viscosity $\alpha=0.1$, and
black hole spin parameter $a=0.95$) are obtained in Fig.3. Dashed
line and solid line are the result from the steady equilibrium
condition and the full calculation by Surman et al. respectively. It
shows that in the inner region of disk(from 20km to 120km), electron
fraction from different methods are conform principally, which
indicates the composition in the disk is in approximate equilibrium
state, but our result is generally smaller than that of Surman et
al. While in outer region of the disk, $Y_e$ deviates from
equilibrium, and this deviation increases with the accretion disk
radius.

Surman and McLaughlin (2004) did not bother with specifying the
radial profile of the temperature and the density of the accretion
disk when calculating the electron fraction as a function of the
radius for model introduced by Popham et al (1999). Here we rewrite
the temperature and the density formulae of Popham et al. (1999)'s
analytical model as
\begin{eqnarray}
T=1.3\times 10^{11}\alpha^{0.2}M_1^{-0.2}R^{-0.3} \rm K\\
 \rho=1.2\times10^{14}\alpha^{-1.3}M_1^{-1.7}R^{-2.55} \dot{M}_1 \rm
 g \, cm^{-3}
\end{eqnarray}
where $M_1$  is  the mass of the accreting black hole in
$M_{\odot}$, and $R$ is the radius in gravitational radius $r_g$
($r_g\equiv GM_1/c^2$, which is equal to 1.4767km for
$M_1=1M_{\odot} $). Since the explicit formulae are given, we adopt
the equilibrium condition of $npe^{\pm}$ gas to obtain some
representative values of $Y_e$ in Fig.4 at the radius larger than
the inner edge (6 gravitational radius) of accretion disk. One can
find from Fig.4 that $Y_e$ have a rapid increase with radius because
both densities and temperatures decrease rapidly when radius
increase and the variation of $Y_e$ is very sensitive to density and
temperature as shown in Fig. 1. For the different accretion rate,
the accretion rate is larger, the $Y_e$ is larger.  This means the
distribution of $Y_e$ along the radius highly depends on the
structure equations of the disk.

\begin{figure*}[htb]
\centering
\includegraphics{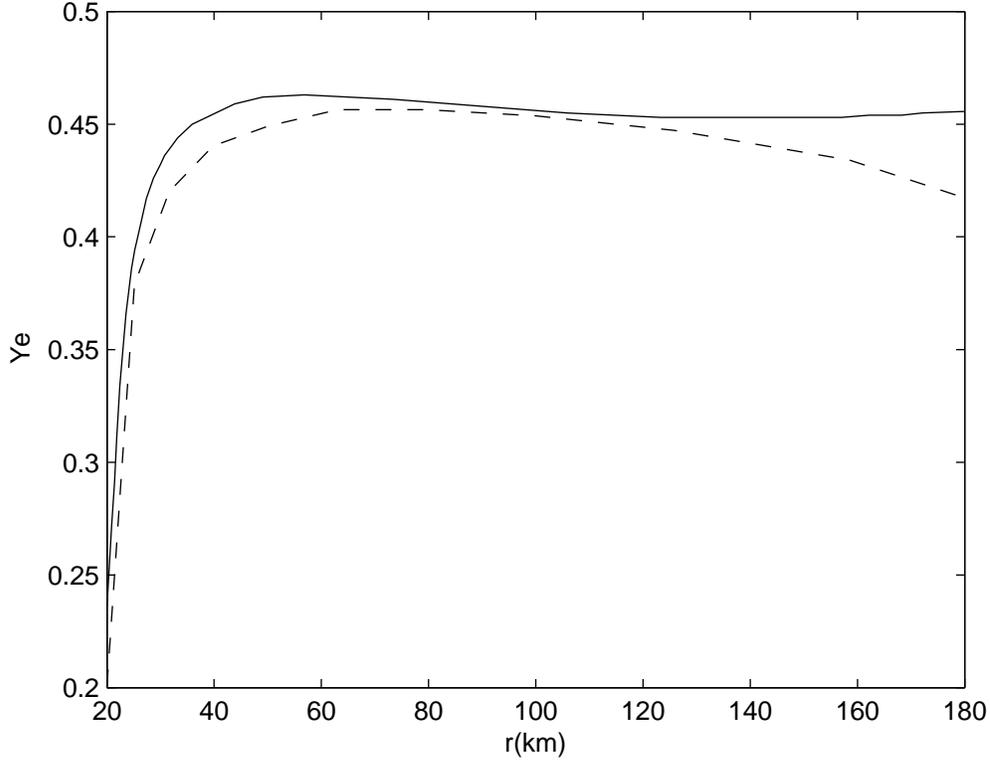} \caption{\label{Fig3}$Y_e$ as a function of accrection disk radius for model
 $\dot{M}=0.1$, alpha viscosity $\alpha=0.1$, and black hole spin parameter a=0.95.
 The dashed line shows $Y_e$ from steady equilibrium condition, while the solid line is the full
  calculation by Surman et al.\citep{Surman2004}}
\end{figure*}

\begin{figure*}[htb]
\centering
\includegraphics{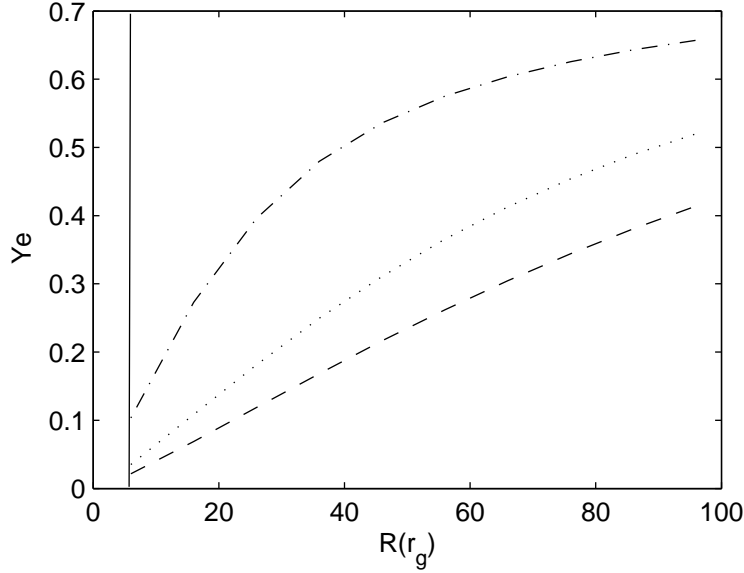} \caption{\label{Fig4} $Y_e$ as a function of accrection disk radius
for the thin disk analytical model($\alpha$=0.1, a=0, $M_1$=3).
Long-dashed line, dotted line and dot-dashed line show $Y_e$ as the
accretion rate $\dot{M}$ =0.01, 0.05, 0.1, respectively. Vertical
solid line denotes the inner boundary of the accretion disk (6
gravitational radius).}
\end{figure*}

\begin{table*}
\centering
 \caption{\small\label{Tab3}steady state chemical equilibrium condition as neutrino opaque for $T=5\times 10^{10}\rm K $.
 The notes are the same as those in Table 1.}
  \vspace{12pt}
\begin{tabular}{c c c c c c c c c c c}
\hline \hline $Y_e$&$\rho$&$\lambda_{e^-p}$&$\lambda_{\nu_e n}$&$\lambda_{e^+n}$&$\lambda_{\bar{\nu}_e p}$&$\lambda_{n}$&$\mu_e^{'}$&$\mu_n^{'}$&$\mu_p^{'}$&$C$\\
$$&$\rm g\,cm^{-3}$&$\rm cm^{-3}\,s^{-1}$&$\rm cm^{-3}\,s^{-1}$&$\rm cm^{-3}\,s^{-1}$&$\rm cm^{-3}\,s^{-1}$&$\rm cm^{-3}\,s^{-1}$&$\rm MeV$&$\rm MeV$&$\rm MeV$&$$\\
\hline

0.10  &       2.32E+11    &   7.92E+37    &   7.89E+37    &   7.88E+36    &   7.56E+36    &   1.23E+31 &   10.18     &   -15.14  &   -24.65  &   1.01    \\
0.20   &       6.17E+10    &   2.03E+37    &   2.01E+37    &   4.17E+36    &   4.01E+36    &   5.87E+30  &   6.69    &   -21.40  &   -27.38  &   1.01    \\
0.30   &       2.46E+10    &   7.35E+36    &   7.26E+36    &   2.48E+36    &   2.39E+36    &   3.10E+30   &   4.37   &   -25.95  &   -29.60  &   1.01    \\
0.40   &       1.05E+10    &   2.75E+36    &   2.71E+36    &   1.40E+36    &   1.35E+36    &   1.52E+30   &   2.48   &   -30.29  &   -32.03  &   1.01    \\
0.50   &       3.40E+09    &   7.56E+35    &   7.40E+35    &   5.59E+35    &   5.43E+35    &   5.17E+29   &   0.75   &   -35.94  &   -35.93  &   1.02    \\
\hline
\end{tabular}
\end{table*}

\subsection{ Case 2. Neutrinos are Opaque}
In the neutrino-opaque and antineutrino-opaque matter, neutrino and
antineutrino will be absorbed by proton and neutron except the
reactions (\ref{2})-(\ref{4}) as following,
 \begin{eqnarray}
 \nu_e +n\rightarrow  e^-+p \label{11},\\
 \bar{\nu}_e+p \rightarrow  e^++n. \label{12}
\end{eqnarray}
 By using sections
$\sigma^{abs}_{\nu_{e}n}=\frac{A}{\pi^2}(E_{\nu_e}+Q)[(E_{\nu_e}+Q)^2-1]^{1/2}(1-f_e)$
and
$\sigma^{abs}_{\bar{\nu}_{e}p}=\frac{A}{\pi^2}(E_{\bar{\nu}_e}-Q)[(E_{\bar{\nu}_e}-Q)^2-1]^{1/2}(1-f_{e^+})$,
we obtain their rates(the natural units system)
 \begin{eqnarray}
\lambda_{\nu_e n}=\frac{A}{2\pi^4}n_n\int_0^{\infty}
(E_{\nu}+Q)[(E_{\nu_e}+Q)^2-1]^{1/2}F(Z+1,E_{\nu_e}+Q)(1-f_e)E^2_{\nu_e}f_{\nu_e}dE_{\nu_e},\\
\lambda_{\bar{\nu}_ep}=\frac{A}{2\pi^4}n_p\int_0^{\infty}
(E_{\bar{\nu}}-Q)[(E_{\bar{\nu}_e}-Q)^2-1]^{1/2}F(-Z+1,E_{\nu_e}-Q)(1-f_{e^+})E^2_{\bar{\nu}_e}f_{\bar{\nu}_e}d\bar{\nu}_e,
\end{eqnarray}
 where $f_{\nu_e}$ and $f_{\bar{\nu}_e}$ are the Fermi-Dirac
distribution function of neutrino and antineutrino.
$f_{\nu_e}=[1+\exp{(\frac{E_{\nu_e}-\mu_{\nu_e}}{kT})}]^{-1}$,
$f_{\bar{\nu}_e}=[1+\exp{(\frac{E_{\bar{\nu}_e}-\mu_{\bar{\nu}_e}}{kT})}]^{-1}$.
The number densities of neutrino and antineutrino are
\begin{equation}
n_{\nu_e}-n_{\bar{\nu}_e}=\frac{4\pi}{h^3}\int
p^2dp\frac{1}{1+\exp{(\frac{E_{\nu_e}-\mu_{\nu_e}}{kT})}}-\frac{4\pi}{h^3}\int
p^2dp\frac{1}{1+\exp{(\frac{E_{\bar{\nu}_e}+\mu_{\nu_e}}{kT})}}.
\end{equation}
When $n_{\nu_e}=n_{\bar{\nu}_e}$, i.e. number density of neutrino is
equal to that of antineutrino,  $\mu_{\nu_e}=\mu_{\bar{\nu}_e}=0$.
In this case, the equilibrium condition Eq.(\ref{5}) becomes
\begin{equation}
 \lambda_{e^{-}p}-\lambda_{\nu_{e}n}=\lambda_{e^+n}-\lambda_{\bar{\nu}_ep}+\lambda_{n}.
 \label{16}
 \end{equation}
One can find from Table \ref{Tab3} that even at $T=5\times
10^{10}$K, $C$ is still approximate to 1. In another word, for a
system with neutrino and antineutrino are opaque and their chemical
potentials are zero, $\mu_n=\mu_p+\mu_e$ is always effective no
matter what temperature is, just as expected.

 \begin{table*}
\centering
 \caption{\small\label{Tab-4}the evolution of initial electron fraction at different steady state chemical equilibrium conditions.
$t$ is the time post bounce, $R_{\nu}$ is the neutrinospheric
radius, $L_n$ is the number luminosity for neutrino and
antineutrino,  $<E_{\nu_e}>$ and $<E_{\bar{\nu}_e}>$ are the average
 energy of neutrino and antineutrino respectively. All parameters above refer to Ref.\citep{Arcones2008}}
  \vspace{12pt}
\begin{tabular}{c c c c c c c c c c}
\hline \hline $t$&$R_{\nu}$&$T$&$L_n$&$<E_{\nu_e}>$&$<E_{\bar{\nu}_e}>$&$\rho$&$Y_e^a$&$Y_e^b$&$C$\\
$s$&$\rm km$&$\rm MeV$&$\rm 10^{56}s^{-1}$&$\rm MeV$&$\rm MeV$&$\rm g cm^{-3}$&$$&$$&$$\\
\hline
2   &   10.55   &   6.34    &   6.05    &   20.71   &   25.64   &   5.50E+11      &   0.113   &   0.084   &   1.39    \\
5   &   9.82    &   5.14    &   3.55    &   17.1    &   22.6    &   1.30E+12       &   0.050    &   0.039   &   1.22    \\
7   &   9.68    &   4.73    &   3.03    &   15.9    &   21.69   &   1.40E+12      &   0.042   &   0.035   &   1.15    \\
10  &   9.59    &   4.37    &   3.06    &   15.05   &   21.86   &   2.00E+12      &   0.029   &   0.028   &   1.03    \\

\hline \label{Tab-4}
\end{tabular}
\end{table*}

\section{Equilibrium Condition of $npe^{\pm}$ Gas with External
Neutrino Flux}

As discussed in Section 2, we only consider that $npe^{\pm}$ gas is
isolated, but for many astrophysical environments, the external
strong neutrino and antineutrino fluxes can not be ignored. These
processes involve some complex and difficult problems that concern
both the neutrino transport and the interactions with nucleons. Here
we discuss the neutrino-driven wind (NDW) from proto-neutron star
(PNS) as a typical example. NDW is regarded as the major site for
the r-process nucleosynthesis according to the observations of
metal-poor-star in the recent years \citep[see
e.g.,][]{Qian2008,Qian2000,Martinez2008}. Since the NDW is firstly
proposed by Duncan et al. in 1986 {\citep{Duncan1986}}, many
detailed analysis for this process have been done by many authors,
including Newtonian and general relativity hydrodynamics and the
other physical inputs, e.g. rotating, magnetic field, termination
shock and so
on\citep{Qian1996,Thompson2003,Metzger2007,Kuroda2008,Thompson2001,Fischer2009}.
A basic scenario of r-process nucleosynsis in the NDW can be simply
described as\citep[see][]{Martinez2008}: soon after the birth of
PNS, lots of neutrinos are emitted from the surface of PNS; because
of
 the photodisintegration of shock wave, the main composition at the surface
 of PNS is proton, neutron, electron and positron (i.e. $npe^{\pm} \rm \,
 gas$);
  in the circumambience of PNS, the main reactions are the neutrino
 or antineutrino's absorption and emitting by nucleons (so called '
 neutrino heat region'); in the further region electron fraction $Y_e$ keeps as a constant and $\alpha$
 particles are combined; above this region, other particles, such as $^{12}\rm C, \,^{9}\rm Be$, are produced till the seed nuclei; abundant neutrinos are captured by seed nuclei in
 succession. The previous researches show that the steady state is a good approximation to the NDW in the first 20
seconds\citep{Thompson2001,Thompson2003,Qian1996,Fischer2009};

 Usually,
 neutron-to-seed ratio, electron fraction, entropy and  expansion timescale are four essential parameters
 for a successful r-element pattern. It is very
 difficult to fulfill all those conditions self-consistently. Electron fraction $Y_e$ is one of
 the most important parameters. Recent research by Wanajo et al. shows that the puzzle
 of the excess of r-element of $A=90$  may be solved  if  $Y_e$  can increase 1-2\%\citep{Wanajo2009}.
 The evolution of $Y_e$ is usually
obtained by solving the differential equation group
 which is related to the EoS, neutrino reaction rates and hydrodynamic frame\citep{Thompson2001}. Initial  $Y_e$ at the origin of wind
 is an important  boundary condition. Considering the neutrinos are emitted
from the neutrino sphere, $Y_e$ at neutrino sphere can be regarded
as the initial $Y_e$ of the wind. For a given model, the initial
$Y_e$ can be determined by making the assumption that the matter in
neutrino sphere is in beta equilibrium\citep{Arcones2008}. To
compare the results with the previous work of Arcones et al.,  we
employs the same PNS model M15-l1-r1
\citep{Arcones2008,Arcones2007}. The model has a baryonic mass of
1.4 $\rm M_{\odot}$, obtained in a spherically symmetric simulation
of the parameterized 15 $\rm M_{\odot}$ supernova explosion model.
Detailed research shows that there are a few $\alpha$ particles will
 appear at the neutrino sphere, but number density of  $\alpha$ particle
 is much smaller than that of proton and neutron, so it is reasonable to
 ignore the  $\alpha$ particle effect on electron fraction, i.e., the matter is regarded as $npe^{\pm}$ gas.
  Simultaneity, although lots of neutrino and
antineutrino are emitted from PNS,  their number densities are
equal, which means $\mu_{\nu_e}=\mu_{\bar{\nu}_e}=0$. Since the
neutrino and antineutrino are transparent to the matter at neutrino
sphere, neutrino produced by reactions (\ref{2})-(\ref{3}) can not
interact with nucleons, but for the neutrino and antineutrino come
from the core region of PNS, absorption reactions (\ref{11}) and
(\ref{12}) are permitted. Their rates are
\begin{equation}\label{eqc}
  \lambda_{\nu_en}=\frac{L_{n,\nu_e}}{4\pi R_{\nu}^2}\sigma_{\nu_e n}^{abs} \rho(1-Y_e)N_A ,
 \end{equation}
 \begin{equation}\label{eqc}
  \lambda_{\bar{\nu}_ep}=\frac{L_{n,\bar{\nu}_e}}{4\pi R_{\nu}^2}\sigma_{\bar{\nu}_e p}^{abs} \rho Y_eN_A ,
 \end{equation}
where $L_{n,\nu}$ and $L_{n,\bar{\nu}_e}$ are the number luminosity
of neutrino and antineutrino respectively, $R_{\nu}$ is the
neutrinospheric radius. Considering too many physical factors (EOS ,
transport equation and so on) will influence the number luminosity
and the neutrino energy, we simply assume the number luminosity and
the energy of neutrino and antineutrino are the same as those in the
wind. Firstly, we obtain the electron fraction by using a general
equilibrium condition
$\lambda_{e^{-}p}-\lambda_{\nu_en}=\lambda_{e^+n}-\lambda_{\bar{\nu}_e
p}+\lambda_{n}$. In other words, if the density and temperature are
fixed for the equilibrium system, the electron fraction is unique.
Then the coefficient $C$ in the chemical potential equilibrium
condition is determined (leftmost column in Table \ref{Tab-4}) . The
results for  model M15-l1-r1 are shown in Table \ref{Tab-4}. $Y_e^a$
is the
 electron fraction for an extreme case $C=1$,  which is adopted in reference \citep{Arcones2008};   $Y_e^b$
is the result in
 which the steady equilibrium condition is valid and the external neutrino flux
 is also considered. We can find $Y_e^b$ is universal smaller than $Y_e^a$,
 which means the external neutrino flux strongly influences the composition of equilibrium system.
 Comparing $Y_e^a$ with $Y_e^b$, one can find that the improved equilibrium condition makes the
electron fraction decrease significantly when the time is less than
5 seconds post bounce. After 5 seconds the electron fractions are
similar to the case $C=1$. Note that it is just a conclusion for the
model M15-l1-r1. Due to the huge difference between the different
models, the results may be quite different for the other models.
More detailed consideration will be done in our further work.
Initial electron fraction is an important boundary condition to
determine the electron fraction of the wind. Since r-process
nucleosynthesis is strongly dependant on the electron fraction, the
accurate electron fraction is useful for the final r-process
nucleosynthesis.

\section{Conclusions}
In this work, we derive the chemical potential equilibrium
conditions  $\mu_n=\mu_p+C\mu_e$ for $npe^{\pm}$ gas at two cases (
with/without external neutrino flux).  Especially in the
neutrino-transparent matter, employing the fitting
Eq.(\ref{fitting}) for the transition from low temperature and high
temperature is a more convenient method than the calculation of
interaction rates as usual. Since chemical potentials are dependant
on three parameters: density, electron fraction and temperature, any
one of those three parameters can be determined if the other two
parameters are given. Although the variation of factor $C$ is
complicated as the external neutrino flux cannot be ignored, one can
obtain the extremum of those parameters assuming the $C=1$ or 2.
Furthermore, our results can be regarded as the reference value for
non-equilibrium sates. Considering the simplicity and the
far-ranging astrophysical environment, the results in this paper is
expected to be used widely in the further relative works.\\

\textbf{ACKNOWLEDGEMENTS}

 The author would like to thank Prof. Yuan Y.-F. for many valuable conversations and help with preparing this manuscript,
 and the referee for his/her constructive suggestions which are helpful to improve this manuscript.

\end{document}